\documentclass[aps,prb,preprint,superscriptaddress,showpacs,floatfix]{revtex4}

\usepackage{graphicx}
\graphicspath{{figs/}}
\begin{document}
\title{Thermal expansion in multiple layers of graphene}
\author{Jin-Wu~Jiang}
    \affiliation{Department of Physics and Centre for Computational Science and Engineering,
             National University of Singapore, Singapore 117542, Republic of Singapore }
\author{Jian-Sheng~Wang}
    \affiliation{Department of Physics and Centre for Computational Science and Engineering,
                 National University of Singapore, Singapore 117542, Republic of Singapore }
\date{\today}
\begin{abstract}
In this file, we apply the nonequilibrium Green's function method to calculate the coefficient of thermal expansion (CTE) in multiple layers of graphene. We focus on the effect from different layer number $N$.

Our main prediction is that: the increase of $N$ can either enhance or weaken CTE, depending on the strength of the substrate interaction $\gamma$. If $\gamma < \epsilon$, where $\epsilon$ is the inter-layer interaction, the CTE will increase with increasing $N$. Otherwise, if $\gamma > \epsilon$, CTE will decrease with increasing $N$.
\end{abstract}

\maketitle
\tableofcontents
\pagebreak

\section{nonequilibrium Green's function method and interaction potential}
We calculate the CTE of the multiple layers of graphene (MLG) using the nonequilibrium Green's function method, which includes all phonon modes automatically and can be applied to all temperatures as it is a pure quantum approach.\cite{JiangJW} In this method, firstly, the averaged vibrational displacement of atom $j$, $\langle u_{j}\rangle$, is calculated through: $\langle u_{j}\rangle=i\hbar G_{j}$, where $G_{j}$ is the one point Green's function and can be calculated from its Feynman diagram expansion in terms of the nonlinear interaction. We consider the nonlinear interaction as,
\begin{eqnarray}
H_{n} & = & \sum_{lmn}\frac{k_{lmn}}{3}u_{l}u_{m}u_{n}+\sum_{opqr}\frac{k_{opqr}}{4}u_{o}u_{p}u_{q}u_{r},
\end{eqnarray}
where $u_{l}$ is the vibrational displacement of atom $j$, multiplied by the square root of its mass. The nonlinear force constant is extracted from ``General Utility Lattice Program" (GULP)\cite{Gale} by the finite difference method. Then the CTE can be calculated explicitly by its definition,
\begin{eqnarray}
\alpha_{j}=\frac{d\langle u_{j} \rangle}{dT}\times\frac{1}{x_{j}},
\end{eqnarray}
where $x_{j}$ is the position of atom $j$ along the expanding direction in the graphene sheet. We can get the final value of CTE by averaging $\alpha_{j}$ over all atoms. To avoid possible boundary effects, we have dropped $10\%$ atoms which are on the two boundaries along the expanding direction. We apply periodic boundary condition in the perpendicular direction. Throughout this file, the $z$ axis is perpendicular to the graphene sheet, and $x$ $y$ axes are lying in the graphene plane.

\begin{figure}[htpb]
  \begin{center}
    \scalebox{1.0}[1.0]{\includegraphics[width=\textwidth]{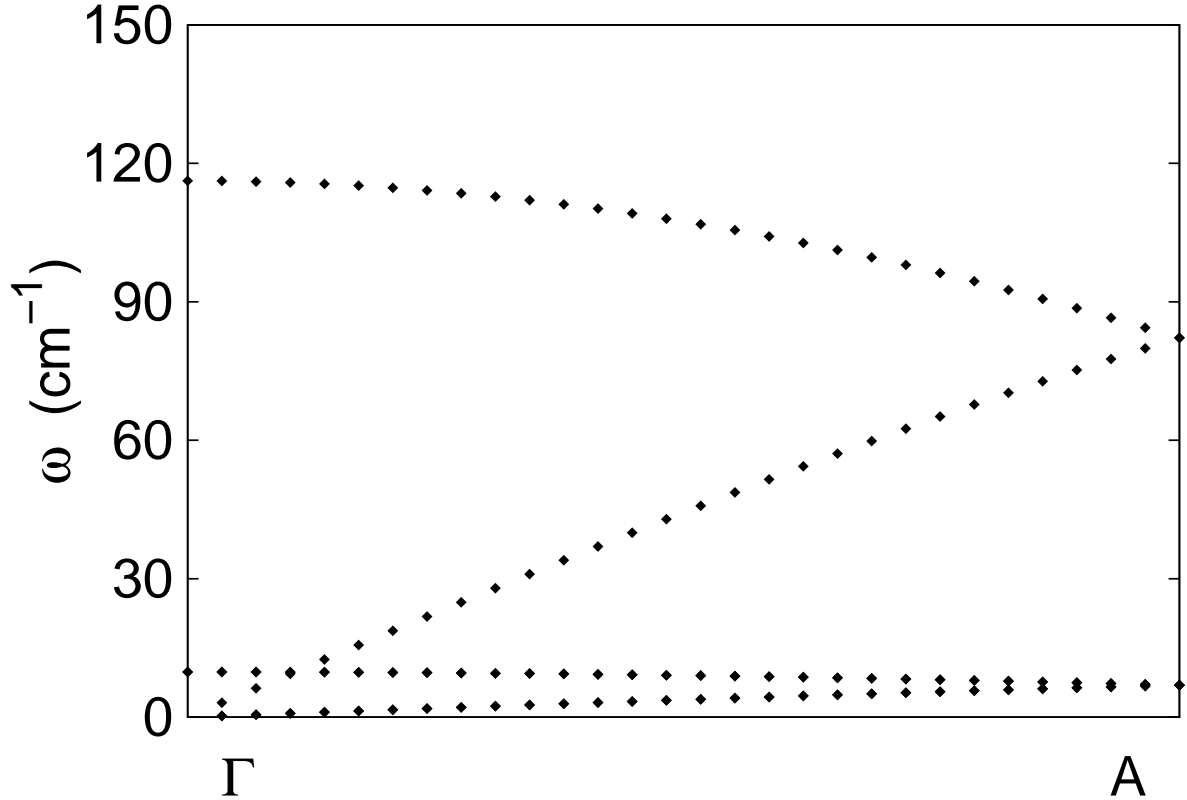}}
  \end{center}
  \caption{Phonon dispersion of three-dimensional graphite along $\Gamma$A direction in the Brillouin zone.}
  \label{fig_phonon_graphite_GA}
\end{figure}
The Tersoff potential is used to describe the intra-layer carbon carbon interactions. Its accuracy has been confirmed by comparison with results from {\it ab initio} density functional theory.\cite{Markussen} For the inter-layer interaction, since the distance between two layers is out of the interaction range of Tersoff, we introduce Lennard-Jones potential, $V(r)=4\epsilon ((\sigma/r)^{12}-(\sigma/r)^{6})$, with $\epsilon=2.5$ mev and $\sigma=3.37$~{\AA}. The value of the length parameter $\sigma$ is fitted to the space between adjacent layers in three-dimensional graphite\cite{SaitoR} as 3.35~{\AA}. The energy parameter $\epsilon$ is fitted to the phonon dispersion along $\Gamma A$ direction in graphite as shown in Fig.~\ref{fig_phonon_graphite_GA}, which is comparable with the experimental results in Ref.~\onlinecite{Nicklow}. The cutoff for Lend-Jones potential is chosen as 20~{\AA}, which is large enough, i.e., all results do not change with further increasing this cutoff value.

The interaction between substrate and carbon atoms in the first graphene layer is simulated by the onsite potential in the $z$ direction\cite{Aizawa}: $V=(\gamma /2) u_{z}^{2}$ with $\gamma$ as the interaction force constant in unit of eV/(\AA$^{2}$u). $u_{z}$ is the vibrational displacement in the $z$ direction multiplied by the square root of mass of the atom. There are three energy scales in the MLG system. The first one is the intra-layer interaction with the force constant in the order of $k_{intra}\approx1$ eV/(\AA$^{2}$u) which is estimated from Ref.~\onlinecite{SaitoR}. In the following context we may simply say that the intra-layer interaction is in the order of 1.0. The second one is the inter-layer interaction with the force constant in the order of $k_{inter}\approx10^{-2}$ eV/(\AA$^{2}$u) which is deduced from the above Lennard-Jones potential. Similarly, without mentioning we would say the inter-layer interaction $\epsilon$ is in the order of $10^{-2}$. The third one is the substrate interaction, which is described by the parameter $\gamma$. When we compare the strength of different interactions, we actually compare their corresponding force constants in the unit of eV/(\AA$^{2}$u).

\section{size and substrate effect}
\begin{figure}[htpb]
  \begin{center}
    \scalebox{1.0}[1.0]{\includegraphics[width=\textwidth]{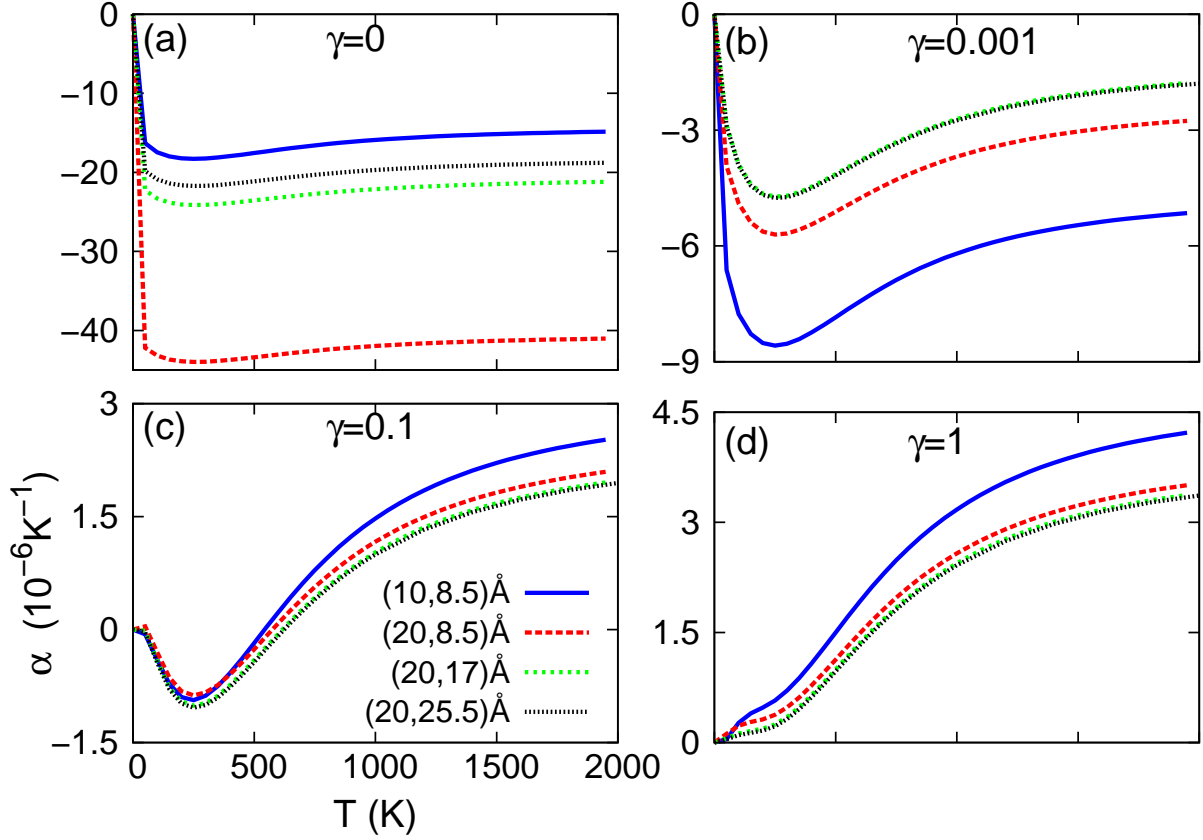}}
  \end{center}
  \caption{CTE v.s. $T$ in single layer graphene sheet with different size, ($length$, $width$), and different substrate interaction $\gamma$.}
  \label{fig_size}
\end{figure}
\begin{figure}[htpb]
  \begin{center}
    \scalebox{1.0}[1.0]{\includegraphics[width=\textwidth]{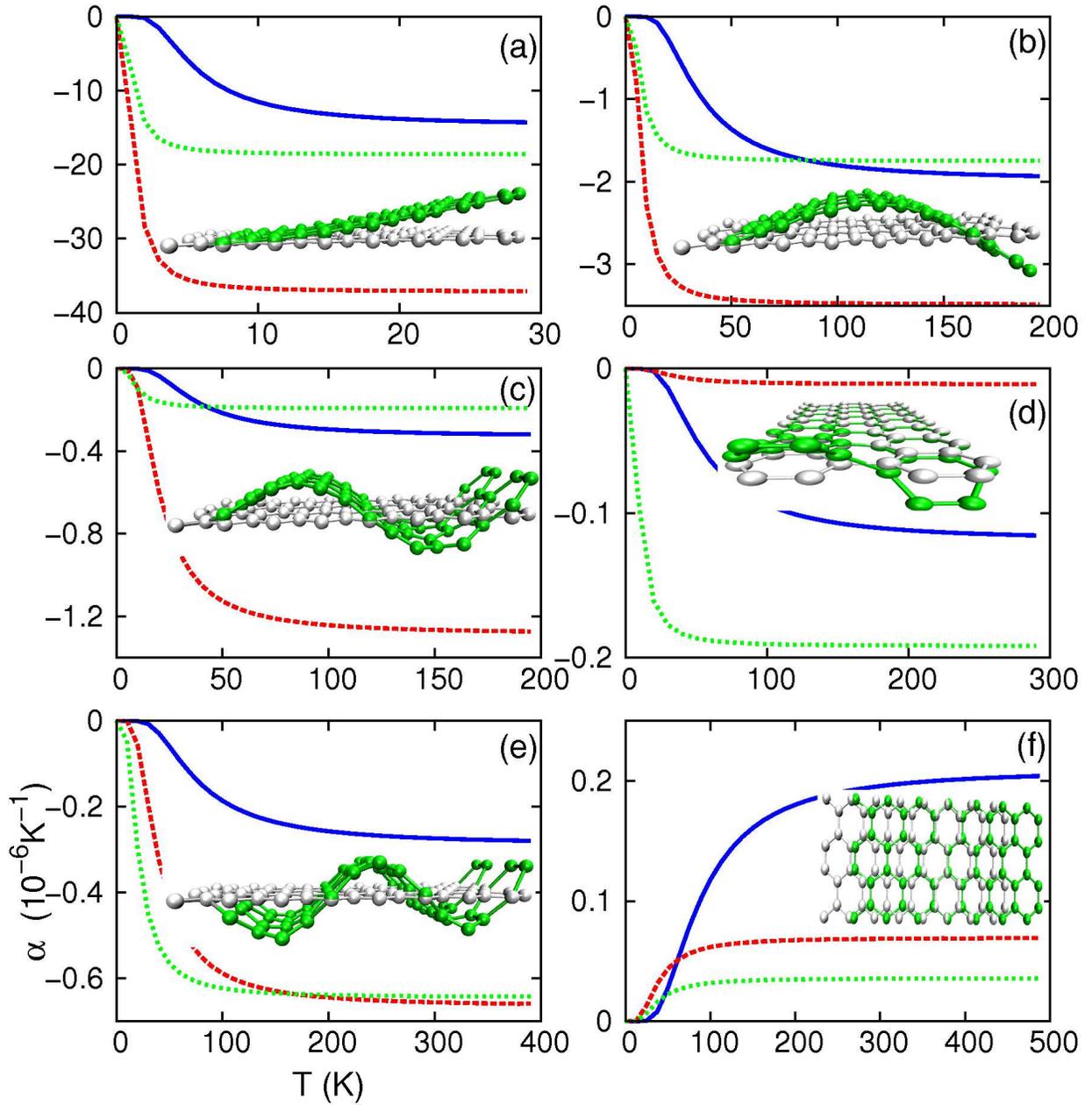}}
  \end{center}
  \caption{Contribution to CTE from six lowest-frequency phonon modes in graphene sheet without substrate interaction. Blue solid line is for graphene with ($length$, $width$)=(10, 8.5)~{\AA}, green dashed line is for (20, 17)~{\AA} and red dotted line is for (20, 8.5)~{\AA}. Insets are the vibrational morphology for the corresponding mode. (a), (b), (c) and (e) are the first four bending modes. (d) is a tearing mode. (f) is the longitudinal vibrational mode.}
  \label{fig_from_each_mode}
\end{figure}
\begin{figure}[htpb]
  \begin{center}
    \scalebox{1.0}[1.0]{\includegraphics[width=\textwidth]{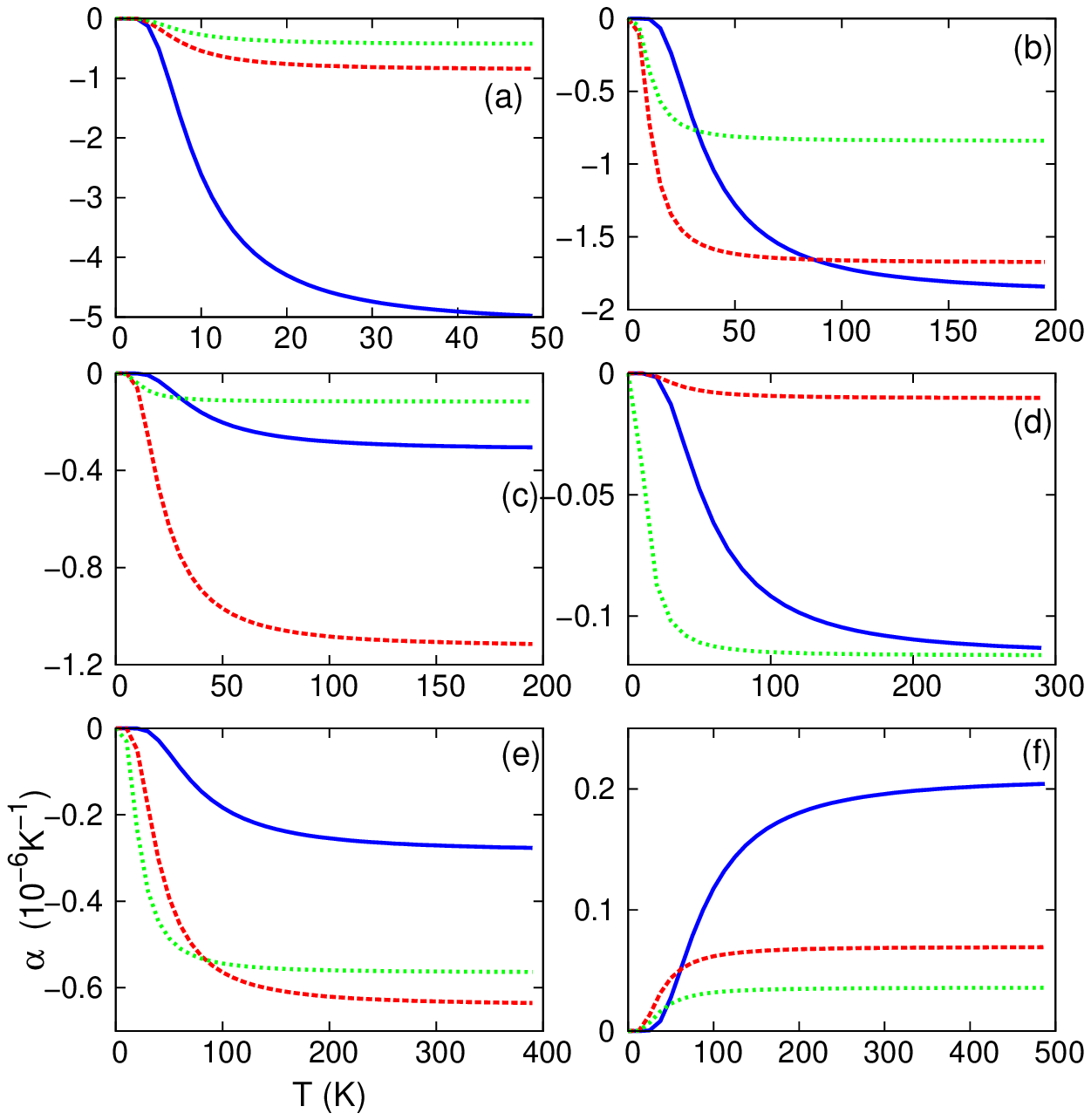}}
  \end{center}
  \caption{Contribution to CTE from six lowest-frequency phonon modes in graphene sheet with substrate interaction $\gamma=0.001$.}
  \label{fig_from_each_mode_gamma_0.001}
\end{figure}

Now we investigate the CTE in single layer graphene sheet with different sizes and substrate interactions. Fig.~\ref{fig_size}~(a) shows that without substrate interaction, the CTE is very sensitive to the size of the system. The value of CTE decreases very quickly with the increase of length. For graphene with length larger than 40~{\AA}, the CTE is a very large negative number, -130 ($10^{-6}$K$^{-1}$), which means that the system with this size is thermally unstable. However, if the graphene sheet is put on a substrate, the situation is quite different. In case of weak substrate interaction, i.e. $\gamma < \epsilon$ as shown in figure (b), the size effect is greatly reduced. Yet it still have some important effect. If the substrate interaction is strong, i.e., $\gamma > \epsilon$ as shown in figure (c), now the size effect is very small. If $\gamma \gg \epsilon$, CTE of all samples are very close to each other at all temperatures and positive in whole temperature range as shown in figure (d).

To understand the size and substrate effect on the CTE, we study the contribution to CTE from different phonon modes. In Fig.~\ref{fig_from_each_mode}, we show contribution from the six lowest-frequency phonon modes. There is no substrate interaction in this figure, i.e., $\gamma=0$. The inset in each panel is the corresponding vibrational morphology of the system in this phonon mode. (a), (b), (c) and (e) are the first four bending modes. (d) is an interesting tearing mode, which may be important in other thermal mechanical process. Among all six phonon modes, we find that the first bending mode shown in (a) dominants the value of CTE in all systems. Its contribution can reach as much as 90$\%$. Due to the morphology of bending movement, it will induce contraction effect in the graphene sheet. So it leads to negative value of CTE. Since all of the first five phonon modes have negative effect on CTE while only the last mode has positive effect, the value of CTE is negative in all samples in case of no substrate interaction. Also, it is easier for longer system to bend,\cite{Krishnan} so the negative effect from the bending mode increases rapidly with increasing length. That is the reason for more serious thermal contraction effect in longer system.

If the substrate interaction is nonzero as shown in Fig.~\ref{fig_from_each_mode_gamma_0.001}, the value of the CTE is enhanced obviously and the difference between different sizes is narrowed. Now the contribution from the second bending mode is also very important. We can also see that the substrate interaction is more important in larger system, while it is less important in smaller system. Because the bending movement in larger graphene is more serious than the smaller system. When the substrate interaction is very strong, the graphene can not bend any more. So all bending modes do not contribute and only the sixth mode makes positive contribution to CTE. As a result, the CTE is positive in whole temperature range, and the difference between systems with different sizes is pretty small.

\section{$N$ effect}

\begin{figure}[htpb]
  \begin{center}
    \scalebox{1.0}[1.0]{\includegraphics[width=\textwidth]{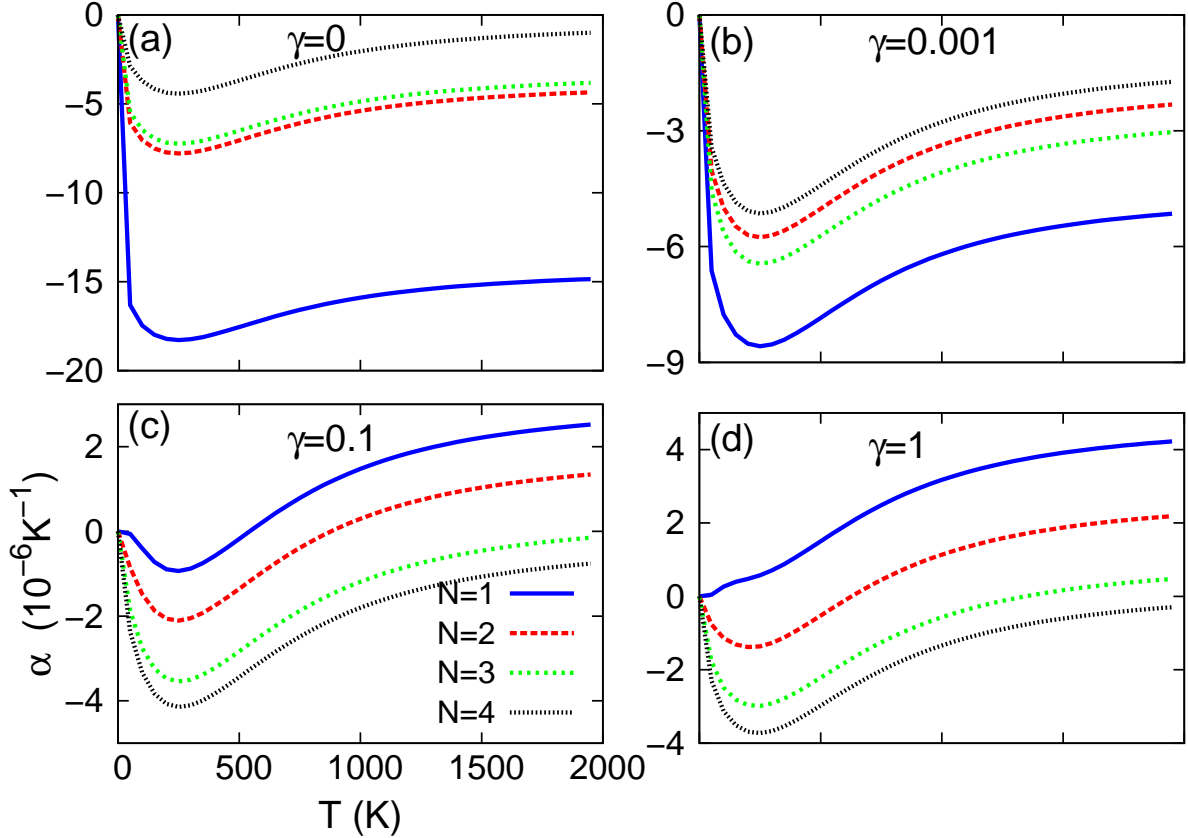}}
  \end{center}
  \caption{CTE in MLG with different layer number $N=1, 2, 3, 4$. The size for each layer is ($length$, $width$)=(10, 8.5)~{\AA}. From (a) to (d) the strength of substrate interaction increases gradually.}
  \label{fig_N}
\end{figure}
\begin{figure}[htpb]
  \begin{center}
    \scalebox{1.0}[1.0]{\includegraphics[width=\textwidth]{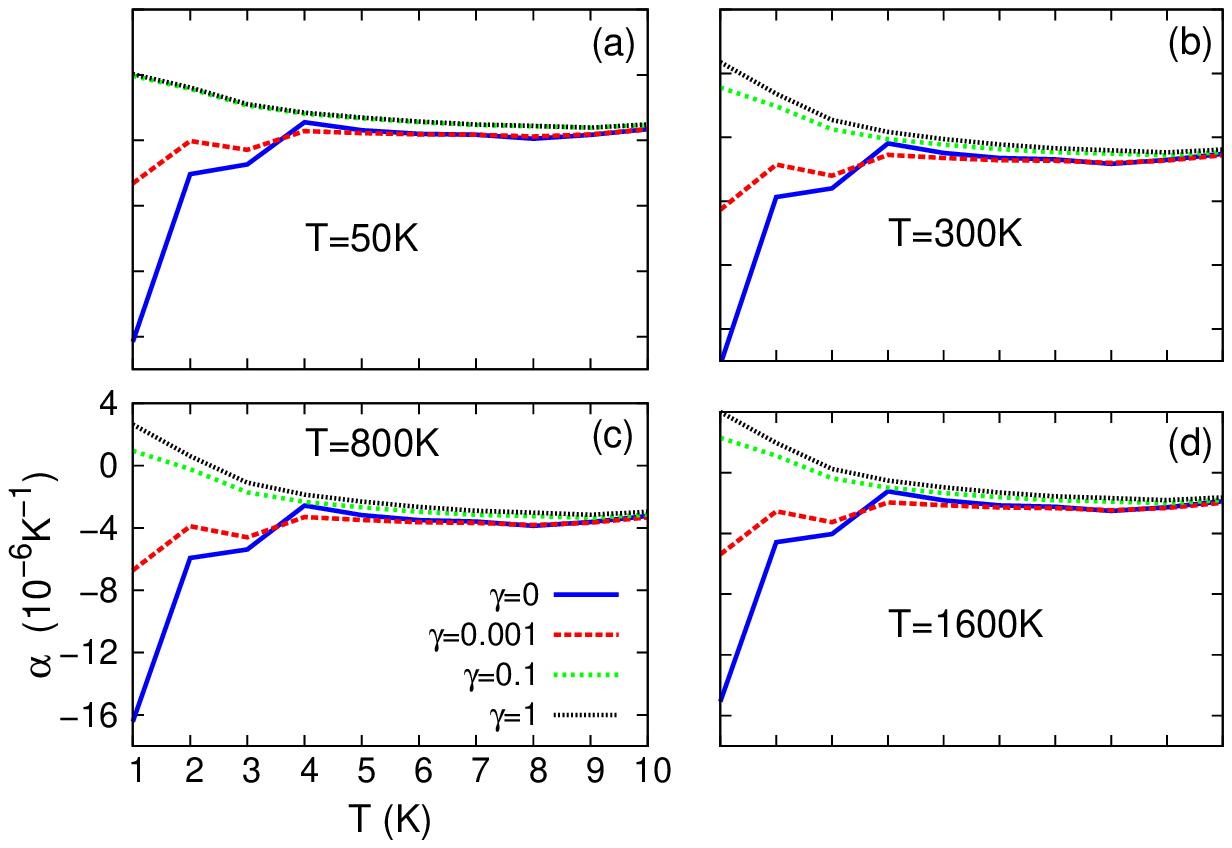}}
  \end{center}
  \caption{CTE in MLG with layer number $N$ from 1 to 10 at different temperatures.}
  \label{fig_more_N}
\end{figure}
Now we study the CTE in MLG with different layer numbers $N$ (from 1 to 4) as shown in Fig.~\ref{fig_N}. In case of no substrate interaction (see panel (a)), CTE increases with the increase of $N$. Typically, when $N$ changes from 1 to 2, the value of CTE shows a big jump. Because the inter-layer interaction makes the bending movement more difficult. So more graphene layers getting together can increase the thermal stability of the MLG if there is no substrate interaction. When the substrate interaction is not very strong, $\gamma < \epsilon$, we can see similar but weaker effect, i.e., the CTE increases gradually with increasing $N$. If the substrate interaction is strong, $\gamma > \epsilon$, we observe opposite phenomenon where the CTE decreases with increasing layer number $N$. At first glance, this is quite strange. However, it is actually physically understandable. If a piece of graphene is transferred onto a substrate with strong interaction, this graphene sheet is very stable. So the bending effect is small, leading to large value of CTE. When the layer number increases from 1 to 2, now the first graphene layer is very stable, but the second layer, which interacts with first layer with $\epsilon$, is not stable. As a result the thermal stability of the whole MLG decreases compared with $N=1$. Similarly, the stability decreases gradually with further increasing $N$. When the substrate interaction is extremely large as shown in (d), we see similar and stronger effect. Fig.~\ref{fig_more_N} shows the layer number dependence of CTE with more values of $N$. We can see that at all temperatures in the figure the CTE increases with increasing $N$ in case of weak or no substrate interaction, and will decrease with increasing $N$ in case of strong substrate interaction. The layer number dependence is more significant before $N=4$. After $N>5$, the difference is very small and the CTE reaches a saturate value independent of the substrate interaction.

\textit{So our theoretical prediction is that: if the substrate interaction is weak or zero ($\gamma < \epsilon$), the value of CTE will increase with the increase of $N$; if the substrate interaction is strong ($\gamma > \epsilon$), the value of CTE will decrease with the increase of $N$.}

\section{possible experiments}
\begin{figure}[htpb]
  \begin{center}
    \scalebox{1.0}[1.0]{\includegraphics[width=\textwidth]{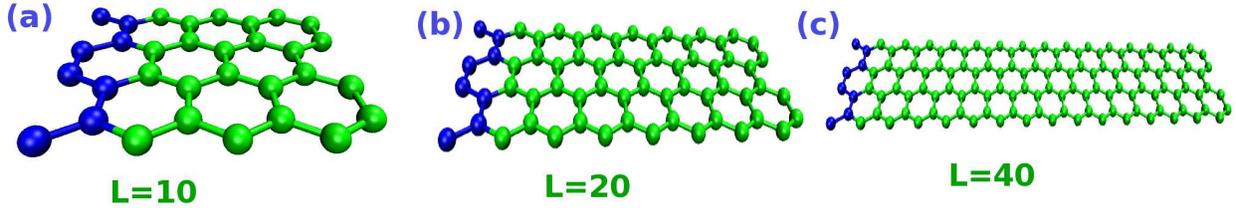}}
  \end{center}
  \caption{Clamp the graphene sample on the left boundary, and increase the size of the graphene. The lengths are 10, 20, 40~{\AA} in three figures. The large sample will be thermally unstable.}
  \label{fig_different_graphene}
\end{figure}
\begin{figure}[htpb]
  \begin{center}
    \scalebox{1.0}[1.0]{\includegraphics[width=\textwidth]{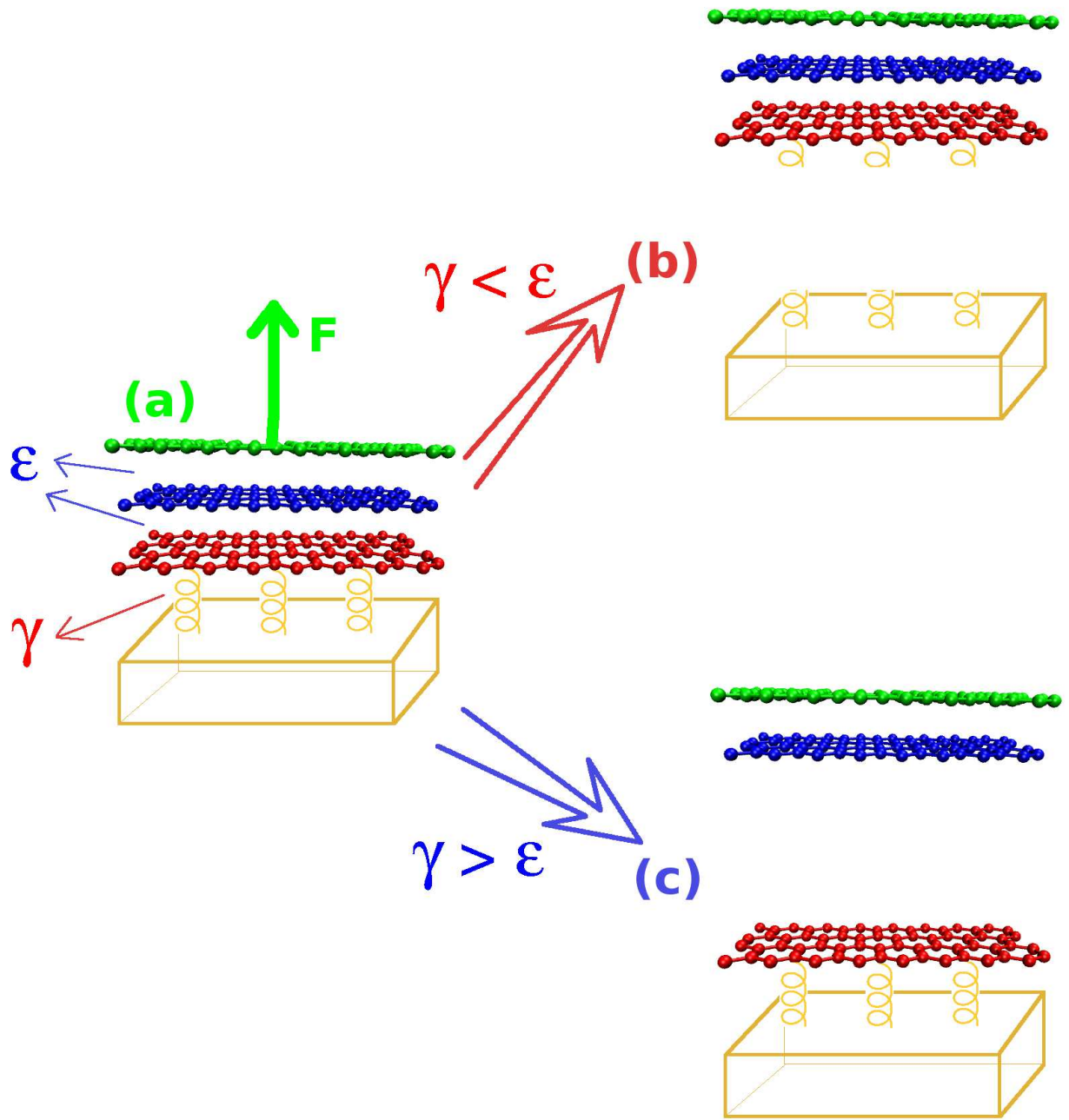}}
  \end{center}
  \caption{Determine the relation between substrate interaction $\gamma$ and the inter-layer interaction $\epsilon$. (a) is a MLG sample with $N=3$ on a substrate. Apply a force $\vec{F}$ to the out-most graphene sheet to pull out the sample. If $\gamma < \epsilon$, the sample will be peeled off the substrate as a whole completely as shown in (b). Otherwise, if $\gamma > \epsilon$, the inner graphene sheet still stick on the substrate while the outer layers of the sample are peeled off as shown in (c).}
  \label{fig_gamma_epsilon}
\end{figure}
In this section, we suppose two possible experiments to confirm our theoretical predictions. The first experiment is to examine the thermal instability of the graphene sheet as displayed in Fig.~\ref{fig_different_graphene}. Single layer graphene sheet with different sizes are picked out and clamped on the left boundary. It is understandable that the graphene sheet will be thermally unstable if the size is very large. According to our theoretical calculation, the typical length is about 40~{\AA}.

The second experiment is to examine the layer number dependence of CTE. The MLG samples with different layer numbers $N$ are transferred onto a substrate and clamped on the left boundary. Then the value of CTE in samples with different $N$ can be measured by Lau's method.\cite{BaoW} We mentioned that although the size of the system in our theoretical study is much smaller than the practical experimental samples, the physical mechanism should be the same in the experiment. According to our prediction, with the increase of $N$, the value of CTE will increase if $\gamma < \epsilon$, or decrease if $\gamma > \epsilon$. So it is very important to determine the relationship between $\gamma$ and $\epsilon$ experimentally. As demonstrated by Fig.~\ref{fig_gamma_epsilon}, this can be done in a very simple way. Figure (a) is a MLG sample on the substrate, where a force $F$ is applied onto the out-most graphene sheet, trying to pull out the MLG sample. If $\gamma < \epsilon$, the whole MLG sample will be peeled off the substrate as shown in (b). On the other hand, if $\gamma > \epsilon$, the first graphene layer will still stick to the substrate, while the other layers are peeled off the substrate shown in (c).

\section*{Acknowledgements}
We thank Wu Gang for helpful comments. The work is supported in part by a Faculty Research Grant of R-144-000-257-112 of NUS.

\end{document}